\documentclass[submission,copyright]{eptcs}
\providecommand{\event}{TFPIE 2019}
\usepackage{breakurl}     
\usepackage{underscore}
\usepackage{amsmath}
\usepackage{listings}
\usepackage{xcolor}
\usepackage{graphicx}
\usepackage[T1]{fontenc} 
\usepackage[utf8x]{inputenc}
\usepackage[title]{appendix}
\usepackage{enumitem}
\usepackage{float}

\usepackage{verbatim}

\usepackage{pgfplots}
\usepackage{inconsolata}
\usepackage[T1]{fontenc}

\usepackage{listings}

\lstdefinelanguage{Clean}{%
    alsoletter={ABCDEFGHIJKLMNOPQRSTUVWXYZabcdefghijklmnopqrstuvwxyz_^`},%
    alsoletter={~!@\#\$\%^\&*-+=?<>:|\\.},%
   morekeywords={generic,implementation,definition,dynamic,module,import,from,while,then,else,where,in,of,case,let,infix,infixr,infixl,class,instance,with,if,derive,otherwise},%
    sensitive=true,%
    morecomment=[l]{//},%
    morecomment=[n]{/*}{*/},%
    morestring=[b]",%
    morestring=[b]',%
    emptylines=1,%
    basicstyle=\small,%
    identifierstyle=\ttfamily,%
    commentstyle=\itshape,%
    keywordstyle={\sffamily\bfseries},%
    stringstyle=\ttfamily,%
    numbers=none,%
    showstringspaces=false,%
    basewidth=0.50em,%
    columns=[c]fixed,%
    keepspaces=true,%
    breaklines=false,%
    tabsize=2,%
    texcl=true,%
    escapeinside={(\#}{\#)},%
    literate=   {\\}{{$\lambda$}}1%
                {A.}{{$\forall$}}1%
                {E.}{{$\exists$}}1%
                {<=}{{$\leq$}}1%
                {>=}{{$\geq$}}1%
                {<>}{{$\neq$}}1%
                {->}{{$\rightarrow$}}2%
                {|->}{{$\mapsto$}}2%
                {=}{{$=$}}1%
                {=>}{{$\Rightarrow$}}2%
                {<-}{{$\leftarrow$}}2%
                {==}{{${=\!\!\!=}$}}2%
                {=->}{{$\Rrightarrow$}}2%
                {==>}{{$\Longrightarrow$}}3%
                {~~}{{$\approx$}}2%
                {~/~}{{$\ncong$}}2%
                {<~}{{$\lesssim$}}2%
                {</~}{{$\lesssim\!\!\!\!\!/\ $}}2%
                {\#}{{$\sharp$}}1%
                {\{|}{{$\!\{\!|$}}1%
                {|\}}{{$\,|\!\}$}}1%
                {<->}{{$\leftrightarrow$}}2%
                {<|>}{{$\updownarrow$}}2%
                {/.\\}{{$\wedge\!\!\!.\,$}}1%
                {/\\}{{$\wedge$}}1%
                {\\./}{{$\vee\!\!\!\cdot\,$}}1%
                {\\.n}{$\backslash$n}2%
                {\\n}{$\backslash$n}2%
                {\\"}{$\backslash$"}2%
                {\\t}{$\backslash$t}2%
                {\\\\t}{$\backslash$t}2%
                {\\\\}{$\backslash$t}1%
                {>>>}{{$>\!\!>\!\!>$}}2%
                {>>==}{{$>\!\!>==$}}5%
                {>>=}{{$>\!\!>=$}}3%
                {>>=.}{{$>\!\!>=\mathbf{.}$}}4%
                {:.}{{$:\!.$}}2%
                {***}{{$*\!\!*\!\!*$}}2%
                {\&\&\&}{{$\&\!\!\&\!\!\&$}}2%
                {<<<}{{$<\!\!<\!\!<$}}2%
                {\{|*|\}}{{$\{\!|\!\!\star\!\!|\!\}$}}3%
}%
\lstdefinestyle{numbers}{numbers=right, stepnumber=1, numberstyle=\tiny, numbersep=-7pt}
\lstnewenvironment{CleanCode}{\lstset{language=Clean}}{}%
\lstnewenvironment{CleanCodeN}{\lstset{language=Clean,style=numbers}}{}%
\lstnewenvironment{CleanCodeNB}{\lstset{language=Clean,style=numbers,frame=single}}{}%
\lstnewenvironment{CleanCodeB}{\lstset{language=Cleanframe=single}}{}%
\newcommand{\Cl}[1]{\lstinline[language=Clean]$#1$}%

\title{How to Increase Interest in Studying\\ Functional Programming \\via Interdisciplinary Application}

\author{Pedro Figueirêdo \and Yuri Kim \and Nghia Le Minh 
       \and Evan Sitt \and Xue Ying \and Vikt\'oria Zs\'ok
\institute{E\"otv\"os Lor\'and University, Faculty of Informatics \\
           Department of Programming Languages and Compilers \\
           Budapest, Hungary}
\email{pedrofigueiredo5206@gmail.com, kyr9412@gmail.com, minhnghia.1999@gmail.com}
\email{Sitt.Evan@gmail.com, xueying19981206@gmail.com, zsv@inf.elte.hu}
}

\def\titlerunning{How to Increase Interest in Studying Functional Programming via Interdisciplinary Application}
\def\authorrunning{P. Figueirêdo, Y. Kim, N. Le Minh, E. Sitt, Y. Xue, V. Zs\'ok}
\begin{document}
\maketitle
\date{}

\begin{abstract}
Functional programming represents a modern tool for applying and implementing software. The state of the art in functional programming reports an increasing number of methodologies in this paradigm. However, extensive interdisciplinary applications are missing. Our goal is to increase student interest in pursuing further studies in functional programming with the use of an application: the ray tracer. We conducted a teaching experience, with positive results and student feedback, described here in this paper.
\end{abstract}

\section{Introduction}
\label{sec:intro}

In recent years, functional programming has become one of the first major programming language courses \cite{course, thompsons, tfpie12}. 
This paper reports the results and the outcomes of an application oriented teaching experience to make a functional programming course more attractive and motivating for first year, first semester students. The experiment was conducted over the span of two semesters from two different academic years. Our goal is to fill the gap between a theoretical teaching approach and application oriented software labs.

Throughout our classes, large amounts of simple exercises were provided to the students so that they could understand recursion, guarded expressions, parameters, and other major functional programming concepts. Past experience has shown that these exercises were helping them; however, they were also keen on applications where they could see functional programming in real world scenarios.

As students come from vastly different backgrounds in high school mathematics and physics, they often struggle to understand the basics and can get lost at the initial stages. This causes the students to become unable to progress in later classes, since the curriculum is fast-paced and iterative.

We considered presenting an especially attractive computer graphics application - ray tracing - wherein, after short introductory presentations, they can extend a partially complete Abstract Data Type (ADT) with new functions and try out well-known concepts, utilized in computer graphics, engineering, medicine, and data visualization applications.

Ray tracing usually is studied extensively in the framework of more advanced courses. However, here we provided first a simple background, then the implementation for the complex parts, leaving it up to the students to extend it by the minor missing properties and operations required to visualize graphical elements. Hence, students were engaged and able to practice various functional programming concepts in the scope of the application and perceive how the acquired language elements would work in practice.

In this paper, we describe our functional programming course structure (section \ref{sec:class}), the motivation of the students via the ray tracing application (section \ref{sec:motivation}), the integration of the application into the course (section \ref{sec:integration}), the presentation and discussion of our results (section \ref{sec:results}), and lastly future work (section \ref{sec:future}) with conclusions (section \ref{sec:conclusion}).

\section{Teaching Introduction to Functional Programming} 
\label{sec:class}
At our university, E\"otv\"os Lor\'and University, an introductory course to functional programming is offered as a compulsory part of the first year first semester curriculum for Computer Science bachelor's students. This is taught concurrently alongside imperative and procedural paradigm courses, such as C/C++ and Python.

In this section, the class structure of our functional programming course (\ref{sec:FPCS}) and the challenges of teaching functional programming to newly admitted students (\ref{sec:chal}) are discussed.

\subsection{Functional Programming Class Structure}
\label{sec:FPCS}
The Functional Programming course is divided into two components. There is a two-hour lecture component and a two-hour practical component. While the lecture component is composed of presentations focused on the theoretical part, the practical component relies on implementation techniques as its primary method of delivering information. The teacher to student ratio is 3:66. In the years prior to our experiment, there was only a two-hour practical session which severely limited the time available, with a teacher to student ratio of 2:83.

There are ten lecture sessions for the course, one per week during the duration of the semester with one week of vacation in the middle of the semester. Each of these sessions focuses on a different aspect of functional programming. They are structured to form a step-wise progression for first-year, first-semester students with no programming knowledge to learn the basics of the functional programming paradigm all the way up to basic abstractions. 

Within the span of the semester, we cover many basic fundamental concepts of functional programming. We start with types, recursion, and function comprehension. Afterwards the course will proceed to lists and list comprehensions, tuples, and records and instances. After the midterm, the lecture proceeds to discuss arrays, abstract data types, and trees. And finally, the course wraps up with classes and input/output.

As the students are first-year, first-semester students, the students are not expected to, nor do most possess, any preliminary knowledge or experience in programming. This means that the lectures not only include functional programming concepts, but must also include general programming concepts and algorithms.

For each lecture session there is a mandatory practical session within the same week. During these practical sessions, the students are given classwork exercises formulated around the theoretical concepts taught during the lecture session. The instructors for the practical sessions help guide the students through these exercises in Clean \cite{clean}. Instructors show how to solve some of those exercises explaining the underlying algorithms and logic which students can use to solve further examples. Additionally, the students are given a short theoretical quiz that helps reinforcing the theoretical concepts that were presented during the prior lecture session along with evaluating their ability to analyze and comprehend code snippets.

\newpage

In addition to the compulsory lecture and practical sessions each week, a consultation session is offered three times per week. During the consultation sessions, teaching assistants are available for the students to consult with. Students are highly encouraged to take advantage of this resource for their homework and general questions.
As the student to instructor ratio is lower during the consultation sessions, the teaching assistants are able to provide more individualized help and explanations to assist each student with understanding the concepts and completing their assignments. Additionally, during the consultations, teaching assistants also help those who are progressing well within the class to learn new methods and techniques in functional programming, along with emphasizing good programming style.

Furthermore, one of our teaching assistants also conducts a more specialized type of consultation via online streaming. These consultations take advantage of current trends in digital media consumption and are held over YouTube. Additional tutorial videos are also made and uploaded to the same channel. This has many advantages over the traditional consultation.

Generally, students were more relaxed as they could attend from the comfort of their dormitory. They had an easier time following the exercises being presented by the teaching assistant as it was on their own screen. Questions were also easier to pose as they could be sent to the chat without disrupting the teaching assistant or drawing attention, and could be addressed by the teaching assistant in an appropriate time. Lastly, with the nature of a digital format, the solved exercises and the recording of the live stream were both easily accessible at any time after the stream, allowing for the students to re-attend the consultation at a later time at their own convenience and allow for students who could not make the scheduled time to have access to it as well.

Each week, the students are assigned a compulsory homework assignment. Each homework assignment has fewer exercises than the classwork assignments, however each exercise is much more sophisticated than the basic classwork exercises. The homework exercises each emphasize and promote a solid understanding of functional programming fundamentals along with the skill for applying and combining them in a close-to-real-life application. Additionally, the difficulty level of the homework is intended to encourage the student to take their time over the span of the week to research and practice the skills required to complete the exercises. It also aims at preparing them for the examination exercises.

Over the course of the semester, students are given two examinations. One is a midterm examination focusing on the first half of the material, while the other is an end term focusing on the second half of the material. Typically the exam questions are selected considering a division regarding difficulties. We try to have basic, intermediate and advanced problems with ratio of 1:2:1. Once instructors and teaching assistants create sample exam questions, they vote what should be included or discarded based on the difficulty criterion and relevance to the course. Students are expected to have a strong understanding of fundamental concepts and the practiced skill to apply them to various applications, and the examination exercises are written with this focus in mind. In the past, it has been observed that the midterm examination also serves as a good measure of progress for the students, and those who take to heart the need for improvement and come to the consultations usually focus on improving on their weaknesses as shown by the midterm. 

By the end of the course, the students are expected to be capable of solving application based tasks via the use of data structures and functional programming techniques. They should also be capable of analyzing code written by themselves or others and understand how to debug or optimize it. Lastly, they are encouraged to develop the skills to translate imperative algorithms into optimized functional algorithms.

\subsection{Challenges}
\label{sec:chal}
As our students are first year first semester computer science students, there are considerable challenges that we have faced in teaching them. Amongst these challenges are their lack of experience with programming and their varying level of mathematical proficiency.

The largest of the challenges is introducing the students to concepts of functional programming concurrently with general concepts of programming. As part of the curriculum provided to the instructors, the students learn the imperative paradigm at the same time, and it becomes increasingly difficult to convey concepts such as immutability. Additionally, as the students learn a total of six programming languages within the span of their first semester, the differences between the other languages, all imperative, and Clean add to the challenge.

The varying levels of mathematical proficiency as a result of different educational backgrounds also adds an additional challenge, as the concepts of functional programming are closely tied to mathematical thinking and theory.

Lastly, but certainly not the least, the majority of the students have not found their "true calling" or passion within computer science yet. As such, the majority are studying the material with no motivation, only doing so because they are required to. It is quite a task to emphasize why they should put effort into learning a paradigm that is very different from imperative programming.

Given that prior to our experiment, the course had only one two-hour practical session per week,  the time allotted to the instructor was too limited to tackle these challenges. With consideration to the repeated requests of the students for an application for functional programming, we decided to find an application to tackle these challenges.

\section{Motivation via Ray Tracing Application}
\label{sec:motivation}
As discussed in Section \ref{sec:chal}, it can be challenging to teach functional programming to first year, first semester students. Therefore, finding an application to demonstrate a use-case of the language could help motivate students to overcome the aforementioned challenges. This section explains the criteria used for selecting an application (\ref{sec:critapp}), provides a background for ray tracing (\ref{sec:rtbg}), and describes the ray tracing application (\ref{sec:rtapp}) proposed by this paper.

\subsection{Criteria for Application Selection}
\label{sec:critapp}
To select an application for motivating students to learn functional programming, it is necessary to decide what constitutes an admissible application for such a goal. For this paper, an admissible application has to fulfill the following requirements:

\begin{itemize}
    \item interdisciplinary,
    \item produces concrete output,
    \item applicable to everyday life of students,
    \item extensible for further improvements,
    \item contains basic and advanced levels of functional programming.
\end{itemize}

An application is considered interdisciplinary if it involves functional programming concepts to solve problems typically encountered in other fields of study. By applying concepts derived from other fields, students can use functional programming to exercise theoretical concepts that are not easily visualized otherwise. Furthermore, the application produces concrete output, permitting the students to see a result (e.g. an image or technical file) by which they can verify the correctness of their solution to the assignment. This provides assistance to students, who are inexperienced with working on larger projects, bolstering their ability to properly analyse their results.

Moreover, an application is applicable to the everyday life of the majority of the students if they can find inspiration from such interactions and realize real-world uses for their work. An extensible application has the ability of adapting to feedback from students as well as adding new features when needed. Finally, to make use of the application throughout the whole semester of teaching, an admissible application should contain basic and advanced concepts to be tested at multiple occasions during the entirety of the teaching period.

\subsection{Background of Ray Tracing}
\label{sec:rtbg}
\begin{figure}[!htb]
    \centering
    \includegraphics[width=100mm]{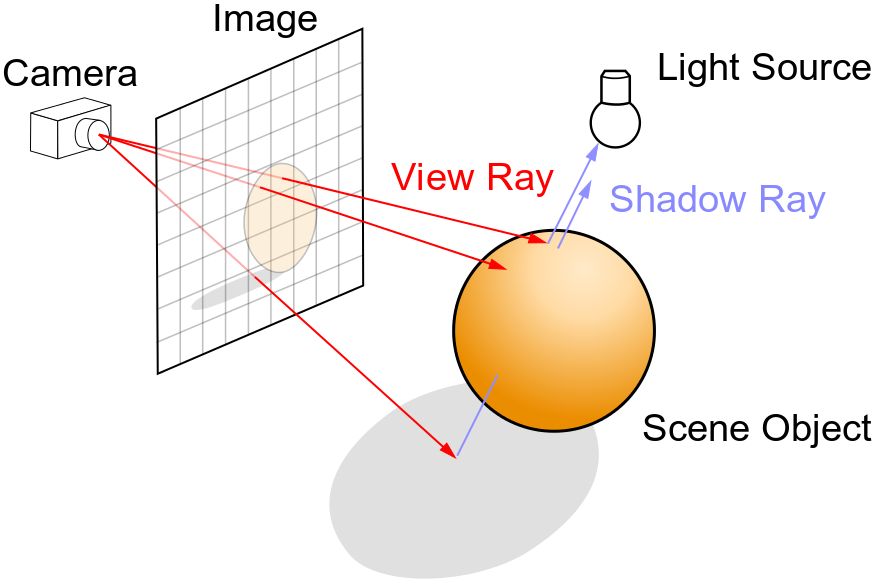}
    \caption{Ray tracing diagram}
    \label{fig:rtdiagram}
\end{figure}

Ray tracing is a technique for image synthesis: creating a 2-D picture of a 3-D world \cite{glassner}. This rendering technique traces the path of light, converting it to pixels in an image plane, as a way of simulating the effects of multiple light rays interacting with primitives in a virtual environment (Figure \ref{fig:rtdiagram}). Its inherently interdisciplinary background, covering multiple topics spanning both Mathematics and Physics, has shown to be a motivational application of theoretical concepts which are considered hard to understand according to undergraduate students \cite{pedro}.

Implementing this algorithm in the context of education in functional programming courses can also demonstrate how well languages in such a paradigm could handle parallel programming \cite{francia, strategy, cefp}. With ray tracing, considered to be an "embarrassingly parallel" \cite{pedro}
 problem, the implementation of such features as well as the understanding of students would be facilitated.

In addition, ray tracing realization encompasses numerous sub-problems, varying in size and complexity, being applicable in different stages of the teaching semester. Beginners could tackle simple algebraic operations while more experienced students can solve ray-triangle intersection tests, for instance. This idea can be applied not only to classwork and homework, but also on tests, thus integrating once minor unconnected tasks into a larger project with a clear and concrete objective: generating images as a result of the ray tracing rendering.

Even though the ray tracing technique is based on various theoretical and lengthy concepts, the core software does not comprise a large project and it does not demand an extensive time investment to be able to display a result. This allows its application by a smaller number of teaching staff. However, the project is not limited to its initial size, being easily expandable and iterative, which can also benefit from a larger staff and experiences from successive semesters.

As it is currently used within a multitude of fields such as engineering, medicine, and data visualization, ray tracing \cite{pedro} can have a motivational appeal to students. Realizing that, by learning functional programming concepts, one can be a part of a ray tracer software project brings meaningful examples of its importance. Having stated that, ray tracing fulfills all the aforementioned criteria (Section \ref{sec:critapp}) for a successful motivating application and was chosen as the motivating application of this paper.

\subsection{Ray Tracing Application Description}
\label{sec:rtapp}
\begin{figure}[!htb]
    \centering
    \includegraphics[width=100mm]{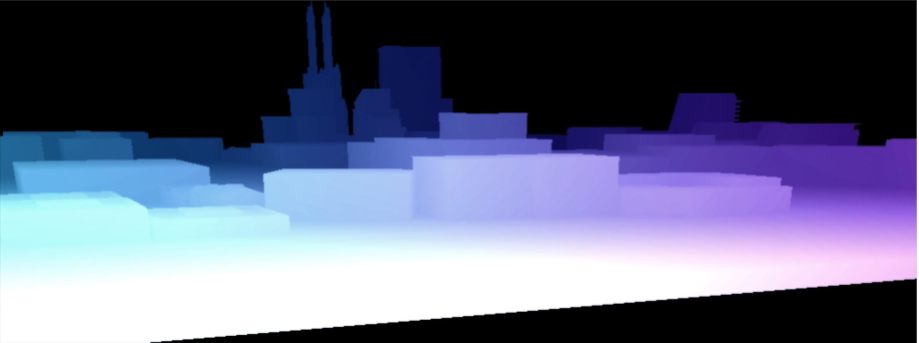}
    \caption{A rendering of a cityscape model}
    \label{fig:cityscape}
\end{figure}

The ray tracing application is a brute-force simplistic ray tracer implemented entirely in the Clean programming language. It reads triangle meshes from files and renders a final scene constructed from ray-triangle intersections. For every pixel of the final image, a ray is shot against the scene made of triangles. The color of the pixel is the color of the triangle closest to the camera, or a predefined background color, if no triangle is intersected. Figure \ref{fig:cityscape} shows a resulting render image from the ray tracing application of a cityscape model consisting of 11,826 triangles.

The application is organized into modules to facilitate its implementation by multiple developers simultaneously. Modular code also increases maintainability, making it easier to test and extend the project on successive semesters of development. There are four groups of modules used by the ray tracing application: \textit{core components}, \textit{camera}, \textit{buffer}, and \textit{raytracer}.

The \textit{core components} group is composed of modules containing \textit{abstract data types (ADTs)} essential to the other parts of the project. It includes the \Cl{VectorOverloading} module, responsible for implementing the \Cl{Vector2} (two-dimensional vector), \Cl{Vector3} (three-dimensional vector), and \Cl{Matrix3} (square matrix of third order) data types and operations involving them. Furthermore, it consists of the \Cl{Triangle}, \Cl{Ray}, and \Cl{IntersectionRecord} modules. Triangles are defined by three \Cl{Vector3} points \textit{a, b, c}, representing their vertices, and a color attribute that stores its \textit{RGB} information in a \Cl{Vector3}. Rays contain a \Cl{Vector3} point of origin, and another \Cl{Vector3} for storing its direction. Lastly, \Cl{IntersectionRecord} ADTs are used for recording intersection information. They store the distance between the intersection point and the ray origin (\Cl{Real}), the intersection point (\Cl{Vector3}), the normal of the intersected object (\Cl{Vector3}), and the color of such object (\Cl{Vector3}).

The \textit{camera} group is composed of two modules: the orthonormal basis (\Cl{ONB}) and the \Cl{PinHoleCamera}. They are responsible for generating a ray for every pixel in the final image to be later intersected against the list of triangles in the scene. \Cl{ONB} \cite{pedro} is an auxiliary module whose task is to generate a matrix used for transforming a ray from the simplified camera space, where it is created, to the world space, where the scene (list of triangles) is located. In performing such a task, it uses \Cl{Vector3} and \Cl{Matrix3} operations defined in the \Cl{VectorOverloading} module. The \Cl{PinHoleCamera} module defines a simple camera without lens which uses perspective projection to depict rendered images. It contains attributes that define the characteristics of the final image (eg. \Cl{resolution_}), and the positioning of the camera (eg. \Cl{position_}, \Cl{up_}, \Cl{lookAt_}).

The \textit{buffer} group consists of only one module of the same name. Its task is to manage file handling: reading from files containing triangle meshes of \textit{obj} type to create a list of \Cl{Triangle} objects from it, and writing the resulting image from the ray tracing into a \textit{ppm} file, which saves \textit{RGB} values for every pixel. It stores the final image as a matrix of \Cl{Vector3} objects, each element representing the color of a pixel.

The \textit{raytracer} group includes an assistance module, which utilizes the M\"oller–Trumbore intersection algorithm for handling ray-triangle intersections (\Cl{RayTriangleIntersector}), and the main module of our application, named \Cl{raytracer}. The M\"oller-Trumbore ray-triangle intersection algorithm is of particular importance as this is typically implemented in the imperative paradigm \cite{moller}. The main module uses all the other groups to generate an image product of ray tracing. It creates a camera with specifications for its positioning and the resolution of the final image. Then, it uses the camera to generate rays for every pixel, according to the previously defined parameters. For every pixel, it intersects the ray with the list of triangles extracted from the input \textit{obj} file. Finally, it saves the result of all intersections into a \textit{ppm} file, showing the renderization of the objects of the input. This particular process is typically implemented in C++. 

\begin{figure}[H]
    \centering
    \includegraphics[width=0.55\textwidth]{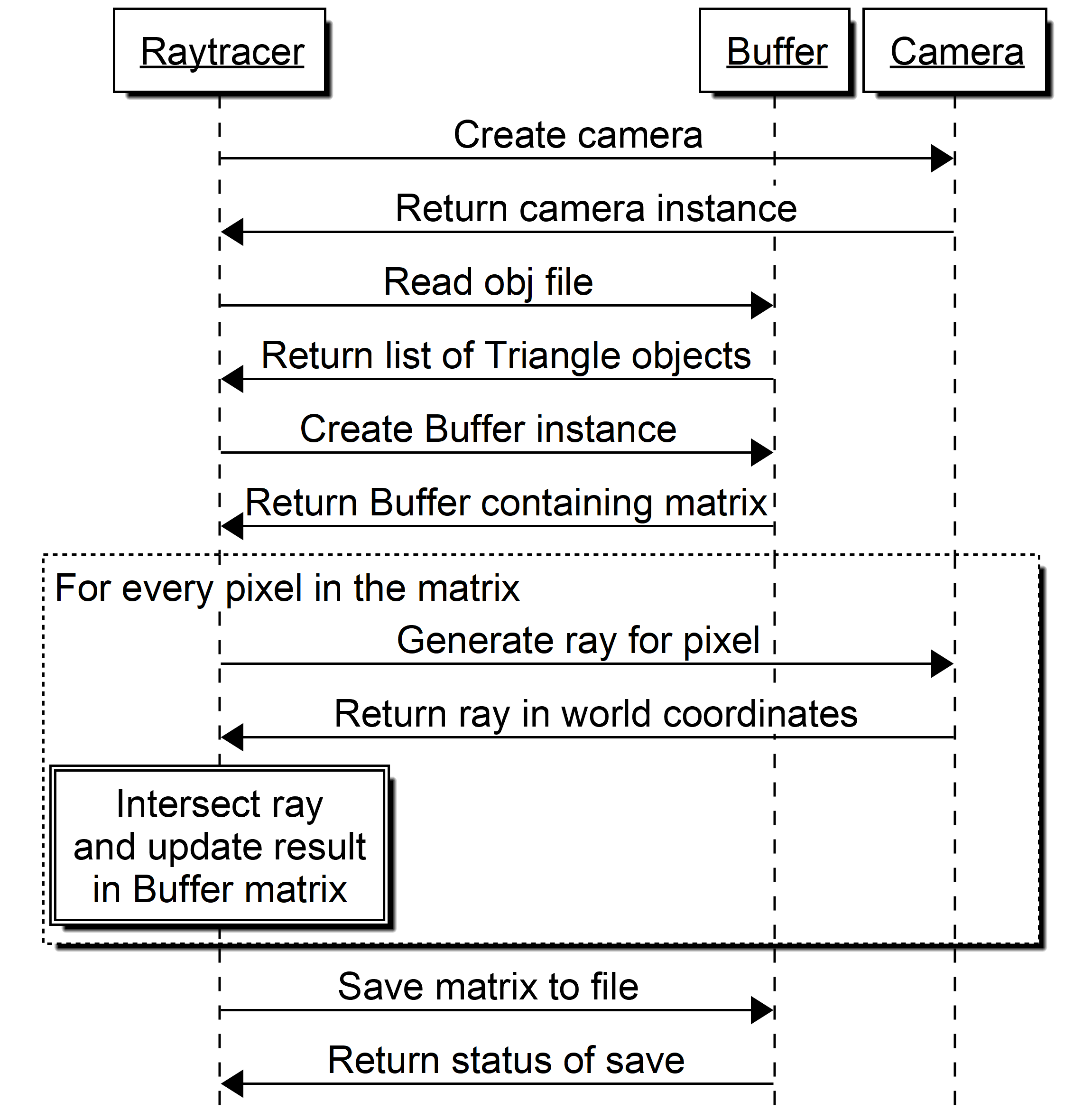}
    \caption{Ray tracing application message sequence chart}
    \label{fig:sigchartrt}
\end{figure}

Figure \ref{fig:sigchartrt} summarizes the interaction between the groups of modules of the ray tracing application in the order of execution through a message sequence chart. The \textit{core components} group is not referred to in the figure, since it consists of a core set of modules used by the other groups.

\section{Integration into the Course}
\label{sec:integration}
Having presented the ray tracing application, its integration into the functional programming course is presented in this section. Firstly, the strategy for integrating the application (\ref{sec:method}) is discussed. Then, the two stages (\ref{sec:expstg}) and two groups (\ref{sec:grpst}) used in the ray tracing application experiment are described. Special integrated exercises (\ref{sec:ex}), with examples (\ref{sec:exqt}), are explained. Finally, ray tracing specific consultations (\ref{sec:cons}) are shown, together with the feedback forms (\ref{sec:fbforms}) used for improving the application on future iterations.

\subsection{Comprehensive Method and Strategy}
\label{sec:method}
Starting from the beginning of the semester, we prepare the students for applying the concepts that they learn in an application. Homework problems are typically simplified versions of real life applications that are encountered in industry. Classwork problems are then respectively tailored to prepare the students for tackling the homework assignments. Additionally, during the lectures and practices it is emphasized when and where certain concepts are used in an application. In this manner, step by step and concept by concept we are able to show the students the reason for the material and its applicability.

As the course approaches the midterm and the concepts start becoming more abstract with the introduction of records and algebraic data types, we started transitioning the students towards the ray tracing project as the ray tracing application is a much more concrete method of demonstrating how these concepts come into play within the context of an application. Additionally, it is easier to demonstrate via an application how the flow of data and information is handled between functions and modules.

\subsection{Stages of the Experiment}
\label{sec:expstg}

To assess the effectiveness of our application on the increase of interest for students, tests were performed in two stages during consecutive semesters. By doing so, not only did we increase the sample size of students, but also performed different ways of integrating the ray tracing project. Applying the experiment over two semesters allowed for testing with different students and comparing the outcome to determine the best integration approach.

For the first stage, we gave the students a brief lecture on the concepts of computer graphics and ray tracing before letting them work on the tasks.
After reviewing the results from the first stage of the experiment, we started the second stage. It involved integrating the material into the course with an independent lecture on computer graphics and ray tracing, with lecture slides and study materials made available to the students. Such integration allowed them more time to better understand the task before having to implement them in their practical sessions and in their homework assignments.

\subsection{Groups of Students}
\label{sec:grpst}
For the first stage of the experiment, as it took place during the spring semester, we had two groups of students. Since the majority of students take the course during the fall semester, the students who took the course during the spring semester were typically students who were retaking the course. These students generally had a harder time with the concepts of functional programming and overall less interest in the material. This formed our first group of students.
To form our second group, we invited students who had completed the course during the fall semester who had demonstrated good performance during the course. This group generally had an interest in the material but had no plans to further pursue the subject proactively.
After the completion of the first stage, we were able to perform another stage of the experiment with the new batch of students in the fall semester. With an increased headcount and with the majority of the students being new students, we had no need to separate them into groups and treated the entirety of the class as a single group (third group).

\subsection{Creation of the special exercises}
\label{sec:ex}
The special exercises were created for the students to put in practice their current knowledge and to use their newly acquired skills in functional programming applications.
For the exercises, we first implemented a finalized program based on Figure \ref{fig:sigchartrt}, which was able to render a red triangle as the default object. Then, the main testing idea was to give the students an incomplete version where the majority of the modules were already provided, only a small chunk of code was removed and then students were asked to implement the deleted parts. The deleted parts could be anything depending on which knowledge (e.g. List Comprehension, Pattern Matching, Tail Recursion, Records, and Abstract Data Type) we wanted to assess. We designed the tests so that, once our students fulfilled these parts, they would be able to produce a red triangle. Typically the students were provided a detailed description of the task, along with the function declaration, both of which are more than sufficient as hints to get the students started on the right path.

In terms of the content of the exercises, since they were the missing parts in the finished version, we created a collection of around ten potential questions, classified into three types:
\begin{enumerate}
    \item Implementing underlying Abstract Data Structures (ADT) (e.g. \Cl{Vector2}, \Cl{Vector3}, \Cl{Matrix3}, and their operators). This class of questions showed our students the usage of records and ADTs.
    \item Implementing the ray-triangles intersection related parts.

    \item Handling input/output: reading points and polygon faces from obj files and writing into \textit{ppm} files.
\end{enumerate}

For composing a test, we first picked from the list above three questions which had an appropriate difficulty for that group based on their previous performance in classwork and homework assignments. The assignments varied according to the groups to better adapt to their experience in programming and, specifically, to the functional programming paradigm. 

In the first and third groups of students, the first assignment involved instance realization (implementing multiple operators) for the complex type \Cl{Vector3}. The second task demanded the dot and cross product 
calculations for the same complex record type. Finally, \Cl{Vector3} normalization was requested as the third assignment. Such test involved easier questions from the exercises of type 1, since they were tailored to first-year first-semester students.

For the second group, the test comprised computing the determinant of a value of type \Cl{Matrix3}, implementing the ray-triangles intersection, and in the end, writing data to \textit{ppm} files (questions of types 1, 2, and 3).

\subsection{Example questions}
\label{sec:exqt}
For those students who found difficulty in grasping the concepts of functional programming, we only asked them to complete instances, functions for \Cl{Vector3}, \Cl{Matrix3} in the \textit{core components} group, which only required a basic understanding of records and ADTs.
\begin{CleanCode}
:: Vector3 a = {x0 :: a, x1 :: a, x2 :: a}

instance +    (Vector3 a) | + a    where + vector0 vector1	= {x0 = (vector0.x0 +
vector1.x0), x1 = (vector0.x1 + vector1.x1), x2 = (vector0.x2 + vector1.x2)}

instance -    (Vector3 a) | - a    where - vector0 vector1	= {x0 = (vector0.x0 -
vector1.x0), x1 = (vector0.x1 - vector1.x1), x2 = (vector0.x2 - vector1.x2)}

\end{CleanCode}

For average students, a wider set of topics could be inquired such as list comprehension, and tail recursion. For instance, with the loop in Figure \ref{fig:sigchartrt}, we demonstrated the usage of tail recursion through the intersection between the ray generated from one pixel and the list of triangles. Furthermore, we asked students to apply the intersection operation to obtain a complete buffer matrix in order to test their understanding of list comprehension.
\begin{CleanCode}
// For one pixel
IntersectRayTriangles :: Ray [Triangle] IntersectionRecord ->
    (Bool, IntersectionRecord)
IntersectRayTriangles ray_ triList initRec =
    IntersectRayTriangles2 ray_ triList (False, initRec)

IntersectRayTriangles2 :: Ray [Triangle] (Bool, IntersectionRecord) ->
    (Bool, IntersectionRecord)
IntersectRayTriangles2 _ [] rec_ = rec_
IntersectRayTriangles2 ray_ [hTri:tTri] (bool_, irec_)
| newIRec.t_ > 0.0 && newIRec.t_ < irec_.t_ && bool_
= IntersectRayTriangles2 ray_ tTri (bool_, irec_)

// For all pixels
integrate resolution triangles = [[integrate2 j i triangles \\i<-[1..x]] \\j<-[y,y-1..1]]
\end{CleanCode}

Additionally, those students who could go further, challenged themselves with the \texttt{Buffer} module. It is especially demanding due to the complexity when using functional programming paradigm in file handling.
\begin{CleanCode}
WriteFile :: *File [[Vector3 Real]] -> *File
WriteFile file img
# file = file <<< "P3\n"
# file = file <<< length (img!!0) <<< ' ' <<< length img <<< '\n' 
# file = file <<< 255 <<< '\n'
# file = foldl (\x y = x <<< y <<< ' ') file img255
= file
\end{CleanCode}

\subsection{Consultations}
\label{sec:cons}
During the consultations that are offered during the course, we help the students with the material. Students are given help with implementing the code, although the assistance provided was limited to hints and suggestions, while implementing the missing pieces was left to the student. We also assisted the students with any mathematical challenges that they were facing in understanding the task. Additionally, we offered to teach more about ray tracing concepts to those who were interested in extra material. 

\subsection{Feedback forms}
\label{sec:fbforms}
We conducted a questionnaire to collect relevant data about how the students 
evaluate introducing application oriented topics with 
its own specific exercises, homework assignments, classwork, assessment quizzes, and lab programming tasks.
The goal was to test qualitative features like 
meaningfulness of the topic for them. By ensuring anonymity of the respondents we maximized the comfort of the students answering. 

\section{Results}
\label{sec:results}

In this section, results comparing the performance of students from both semesters according to classwork (\ref{sec:rescw}), homework (\ref{sec:reshw}), and feedback forms (\ref{sec:resff}) are presented and discussed.

\subsection{Classwork}
\label{sec:rescw}
Since classwork is divided into a theoretical and practical part, two independent analyses were made to compare the integration of the ray tracing application between semesters. Relative metrics were considered to counter the difference in the number of students participating across the semesters. In the first stage, only 8 students participated against 55 in the most recent teaching period.

The performance for the theoretical part of the classwork (short quiz applied in the beginning of the classes) was consistent for both semesters. There was a one percent increase in the mean of performance of students, going from 63\%, in the first stage, to 64\% in the current one. However, two aspects were different in the two applications of the quizzes: the time limit and the phase of the course.

Students had ten minutes to finish a quiz in the first stage whereas pupils from the second stage had only eight (20\% decrease). Furthermore, the test was applied at the end of the first stage, evaluating students that had already seen all the topics in the curriculum and were familiar with them. For students involved in the second stage of the experiment, the quiz of same size and difficulty was taken in the middle of the semester. Therefore, although current students had less time to finish the theoretical part of the classwork and it was applied at a much earlier stage in the course, they maintained the same performance as compared to their predecessors.

In the practical part of the classwork, the same questions were used for groups from each semester. Results displayed in Figure \ref{hist:cw} show a significant improvement in the ability of students from the second stage in solving ray tracing related tasks, when compared to the ones from the first stage.

\pgfplotsset{width=11cm,compat=1.8}
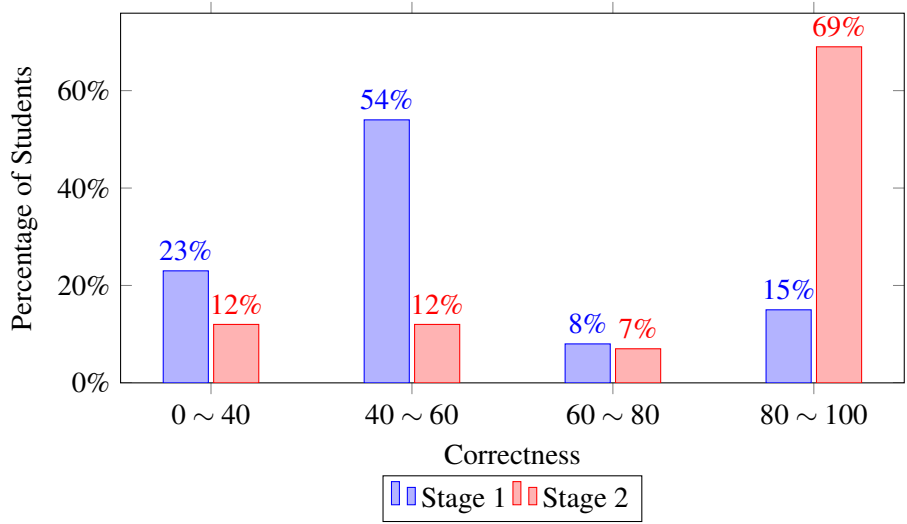
\begin{figure}[H]
\centering
\begin{tikzpicture}
\begin{axis}[
    ybar,
    enlarge x limits=0.15,
    ymin=0,
    legend style={at={(0.5,-0.25)},
      anchor=north,legend columns=-1},
    ylabel={Percentage of Students},
    xlabel={Correctness},
    bar width=0.6cm,
    symbolic x coords={$0 \sim 40$,$40 \sim 60$,$60 \sim 80$,$80 \sim 100$},
    xtick=data,
    yticklabel={\pgfmathparse{\tick}\pgfmathprintnumber{\pgfmathresult}\%},
    nodes near coords = {\pgfmathprintnumber\pgfplotspointmeta\%},
    nodes near coords align={vertical},
    width=12cm, height=6.5cm
    ]
\addplot coordinates {($0 \sim 40$,23) ($40 \sim 60$,54)  ($60 \sim 80$,8) ($80 \sim 100$,15)};
\addplot  coordinates {($0 \sim 40$,12) ($40 \sim 60$,12)  ($60 \sim 80$,7) ($80 \sim 100$,69)};
\legend{Stage 1, Stage 2}
\end{axis}
\end{tikzpicture}
\caption{Classwork Comparison}
\label{hist:cw}
\end{figure}

Figure \ref{hist:cw} shows that 69\% of the students from the second stage achieved a score equivalent to the maximum grade (Excellent, above 80\%) in the Hungarian education system, more than four times the numbers from the first stage. Moreover, the number of students failing (0-40 points) the practical part of the classwork decreased by roughly 50\%.

It is important to state that, as in the theoretical counterpart, the practical part of the classwork was applied in the second stage at a much earlier point than in the first stage. In addition, the time limit for the practical part was the same in both semesters. These two conditions make the aforementioned increase in performance even more significant.

\subsection{Homework}
\label{sec:reshw}

The analysis of homework scores is similar to the one done for the practical part of the classwork: students were given the same time frame to solve the problems of equivalent difficulty and length. However, no ray tracing related homework was applied in the first stage and, thus, no direct comparison could be made to the ray tracing homework done in the second stage. This happened since the limit of the numbers of homework assignments for the first stage had already been achieved when ray tracing concepts were presented at the end of the semester.

Be that as it may, a relevant comparison can be made to homework in which the same functional programming topics were included, and the problems count as well as the time limit for solving them are the same. Figure \ref{hist:hw} shows a comparison between an equivalent homework from the first stage and the ray tracing related one administered in the most recent period of study.

\pgfplotsset{width=11cm,compat=1.8}
\begin{figure}[H]
\centering
\begin{tikzpicture}
\begin{axis}[
    ybar,
    enlarge x limits=0.15,
    ymin=0,
    legend style={at={(0.5,-0.25)},
      anchor=north,legend columns=-1},
    ylabel={Percentage of Students},
    xlabel={Correctness},
    bar width=0.6cm,
    symbolic x coords={$0 \sim 40$,$40 \sim 60$,$60 \sim 80$,$80 \sim 100$},
    xtick=data,
    yticklabel={\pgfmathparse{\tick}\pgfmathprintnumber{\pgfmathresult}\%},
    nodes near coords = {\pgfmathprintnumber\pgfplotspointmeta\%},
    nodes near coords align={vertical},
    width=12cm, height=6.5cm
    ]
\addplot coordinates {($0 \sim 40$,30) ($40 \sim 60$,0)  ($60 \sim 80$,20) ($80 \sim 100$,50)};
\addplot coordinates {($0 \sim 40$,13) ($40 \sim 60$,0)  ($60 \sim 80$,4) ($80 \sim 100$,83)};
\legend{Stage 1, Stage 2}
\end{axis}
\end{tikzpicture}
\caption{Homework Comparison}
\label{hist:hw}
\end{figure}
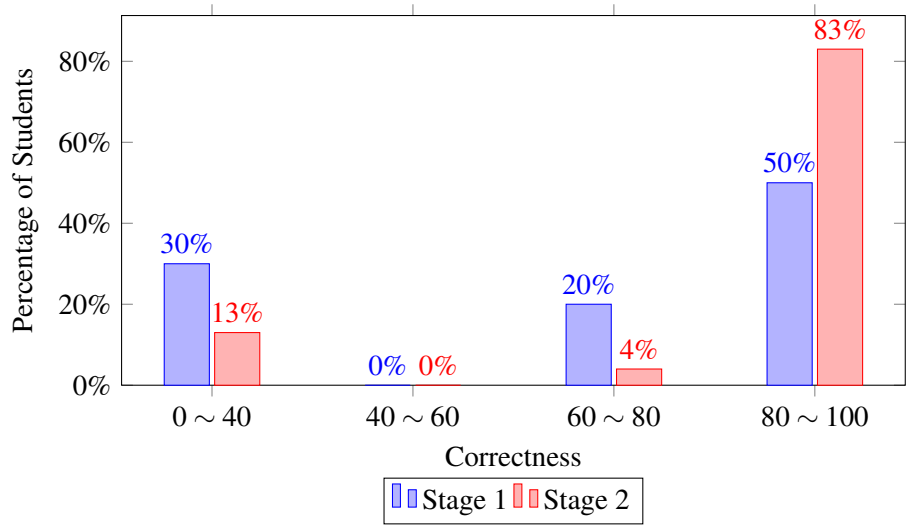

Figure \ref{hist:hw} shows that the ray tracing integration into the homework improved the correctness of the assignments of students in a meaningful way. The greatest increase is seen in the range of 80 to 100 points, going from 50\% to 83\%. Additionally, there has been a decrease of more than half of the students who failed the assignment (0-40 points), even with more challenging problems for the homework applied in the second stage (Section \ref{sec:FPCS}).

\subsection{Feedback forms}
\label{sec:resff}
Feedback forms were collected and analysed identically for both stages of the experiment. The questions are listed below. Respondents could reply with either "Disagree", "Agree", or "Neutral". Figures \ref{hist:fbprev} and \ref{hist:fbcurr} show the responses of the students from stage 1 and stage 2 respectively.

\begin{itemize}
\setlength{\itemindent}{-.2in}
\setlength{\itemsep}{-1.1 pt}
\item Q1: Ray tracing application via functional programming is relevant to my interests.
\item Q2: Ray tracing implementation in function programming is sufficiently difficult for me.
\item Q3: Applying functional programming as part of a larger project helps with understanding functional programming concepts.
\item Q4: I enjoyed implementing the code and was motivated to do so.
\item Q5: Ray tracing has helped my curiosity about functional programming.
\item Q6: Integration of ray tracing into the curriculum helps demonstrate functional programming possibilities.
\item Q7: Enough background was provided to allow me to understand the exercises.
\item Q8: The exercises assigned to me were challenging.
\item Q9: I would be interested in taking an application based advanced Functional Programming course.
\end{itemize}

\pgfplotsset{width=11cm,compat=1.6}
\begin{figure}[H]
\centering
\begin{tikzpicture}
\begin{axis}[
    ybar,
    enlarge x limits=0.15,
    ymin=0,
    legend style={at={(0.5,-0.15)},
      anchor=north,legend columns=-1},
     ylabel={\#Votes},
    symbolic x coords={Q1,Q2,Q3,Q4,Q5,Q7,Q8,Q9},
    xtick=data,
    nodes near coords,
    nodes near coords align={vertical},
    width=0.9\textwidth, height=6.5cm
    ]
\addplot [fill=red!80!white]  coordinates {(Q1,3) (Q2,3)  (Q3,0) (Q4,0) (Q5,1)  (Q7,2) (Q8,3)  (Q9,2)};
\addplot [fill=yellow!50!pink] coordinates {(Q1,2) (Q2,6)  (Q3,2) (Q4,5) (Q5,2)   (Q7,4)  (Q8,3)  (Q9,3)};
\addplot [fill=yellow!50!green] coordinates {(Q1,13) (Q2,9) (Q3,16) (Q4,13) (Q5,15)  (Q7,12)  (Q8,12) (Q9,13)};
\legend{Disagree,Neutral,Agree}
\end{axis}
\end{tikzpicture}
\caption{Stage 1}
\label{hist:fbprev}
\end{figure}

\pgfplotsset{width=11cm,compat=1.6}
\begin{figure}[H]
\centering
\begin{tikzpicture}
\begin{axis}[
    ybar,
    enlarge x limits=0.15,
    ymin=0,
    legend style={at={(0.5,-0.15)},
      anchor=north,legend columns=-1},
    ylabel={\#Votes},
    symbolic x coords={Q1,Q2,Q3,Q4,Q5,Q6,Q7,Q8,Q9},
    xtick=data,
    nodes near coords,
    nodes near coords align={vertical},
    width=1.0\textwidth, height=6.5cm
    ]
\addplot [fill=red!80!white] coordinates {(Q1,8) (Q2,3)  (Q3,4) (Q4,5) (Q5,10) (Q6,4) (Q7,14) (Q8,1)  (Q9,4)};
\addplot [fill=yellow!50!pink] coordinates {(Q1,6) (Q2,3)  (Q3,8) (Q4,9) (Q5,5)  (Q6,8) (Q7,3)  (Q8,8)  (Q9,8)};
\addplot [fill=yellow!50!green] coordinates {(Q1,6) (Q2,14) (Q3,8) (Q4,6) (Q5,5)  (Q6,8) (Q7,3)  (Q8,11) (Q9,8)};
\legend{Disagree,Neutral,Agree}
\end{axis}
\end{tikzpicture}
\caption{Stage 2}
\label{hist:fbcurr}
\end{figure}

The histograms in Figures \ref{hist:fbprev} and \ref{hist:fbcurr} show us several points:
\begin{enumerate}
\item The majority of students find it sufficiently difficult implementing ray tracing in functional programming.
\item In the second stage, most of the students believe that they don't get enough background to understand the exercises.
\item There are more students who are interested in taking an application based advanced Functional Programming course than students who don't.
\item Most students approve the application of functional programming as part of a larger project. They defend the position that it helps with understanding functional programming concepts.
\end{enumerate}

We received in-person feedback from eight of the students in the first stage of the experiment, demonstrating interest in further pursuing learning and developing applications in functional programming. This is double the amount of students who expressed the same interest in the previous year, before the experiment. As of the time of writing, the students are now in progress of doing a research project using Clean to develop a sound synthesizer application (\ref{sec:futureapp}).

The feedback questionnaire presented results indicating that students find functional programming to be better taught when applied to larger projects. It also showed that they enjoyed implementing the ray tracing related exercises, and it has piqued their curiosity towards the functional paradigm, inspiring ideas for new applications (Figure \ref{hist:fbprev}).

It is noticeable by the feedback data that students who perform excellently on the subject of functional programming are even more interested by its applications when compared to regular alumni. This shows that applying functional programming to ray tracing not only piques the interest of regular students, but also provides new challenges to keep excellent students motivated (Figure \ref{hist:fbprev}).

However, compared to the first test groups, the second test groups expressed frustration at the difficulty of the tasks and felt that they did not have enough background or proficiency to tackle the tasks (Figure \ref{hist:fbcurr}). This is likely due to the difference in exercises provided to the students between the first stage and the second stage.

A possible reason for the difference in the two stages perhaps is due to our integration. In the first stage, the experiment was conducted near the end of the semester. As such, the students already had more experience in the course and the tasks given to them were less demanding. With the integration into the course from the second stage, the students were given more demanding tasks at an earlier point in the course. This was most likely the cause of their feelings of being overwhelmed by the material and tasks.

All in all, every group showed improvement in their understanding of functional programming and a marked increase in their interest in further pursuing related projects and advanced courseworks.

\section{Future Work}
\label{sec:future}

With the benefit of hindsight on our experiment, we found multiple points of possible improvement and extension for the future. In this section, ray tracing application improvements (\ref{sec:rtimprov}), the integration of any application into the functional programming course (\ref{sec:appint}), and the development of other applications (\ref{sec:futureapp}) are discussed.

\subsection{Ray Tracing Application Improvements}
\label{sec:rtimprov}
For the ray tracing application, there are significant areas for improvement and expansion. One immediate area for improvement would be to refactor the code for easier readability, better comprehension, and increased efficiency. Additionally, we can add more features to extend the application. One notable feature would be the use of texture files. Another further addition would be the use of multiple lighting sources or lighting types.
\subsection{Application Integration}
\label{sec:appint}
For better integration of the application into the curriculum, we think it would be beneficial to do some improvements. One would be the creation of a web based interactive compiler that would allow the students to solve the exercises, compile, and instantly see their results in a controlled environment akin to the HackerRank model \cite{hackerrank}.

\subsection{Other Applications}
\label{sec:futureapp}
In the future, other applications involving functional programming should also be explored. Currently the team is working on a new application involving sound synthesis, and hopefully with more interest from the students a wider variety of applications can be explored. This in turn can cultivate interests in more students and motivate them to further pursue functional programming.

\section{Conclusion}
\label{sec:conclusion}

Through our experience in teaching functional programming within the given curriculum constraints, we came up against multiple challenges in cultivating an interest in our students and motivating them to further pursue functional programming. We decided upon utilizing an application programmed entirely in the functional paradigm, which inspired students in many ways. They were excited and more self-motivated for learning functional programming and had increased engagement when compared to earlier practical sessions. Since it is widely used in multiple fields and uses multiple functional programming concepts, ray tracing was an attractive application overall for the students. The questions related to ray tracing suited the entirety of the semester, thus being a valuable addition to the curriculum. 

\section*{Acknowledgement}
This work was supported by the European Union, co-financed by the European Social Fund, grant. no \textbf{EFOP-3.6.3-VEKOP-16-2017-00002}. We would also like to extend our eternal gratitude to the students taking part in the teaching experiments and to the demonstrators helping in the preparation and carrying out of the survey.

\nocite{*}
\bibliographystyle{eptcs}
\providecommand{\urlalt}[2]{\href{#1}{#2}}
\providecommand{\doi}[1]{doi: \urlalt{http://dx.doi.org/#1}{#1}}

\newpage
\begin{appendices}

\section{Implementation}
\label{sec:appendixImpl}
\subsection{VectorOverloading}
\begin{CleanCode}
:: Vector3 a = {x0 :: a, x1 :: a, x2 :: a}

instance ==   (Vector3 a) | == a
instance zero (Vector3 a) | zero a
instance one  (Vector3 a) | one a
instance ~    (Vector3 a) | ~ a
instance +    (Vector3 a) | + a
instance -    (Vector3 a) | - a
instance *    (Vector3 a) | * a
instance /    (Vector3 a) | / a

Vec3dotProduct :: (Vector3 a) (Vector3 a) -> a  | *,+ a
Vec3crossProduct ::  (Vector3 a) (Vector3 a) ->  (Vector3 a) | *,-a
Vec3normalize :: (Vector3 a) ->  (Vector3 a) | sqrt,*,+,/ a

:: Vector2 a = {v0 :: a, v1 :: a}

instance ==   (Vector2 a) | == a
instance zero (Vector2 a) | zero a
instance one  (Vector2 a) | one a
instance ~    (Vector2 a) | ~ a
instance +    (Vector2 a) | + a
instance -    (Vector2 a) | - a
instance *    (Vector2 a) | * a
instance /    (Vector2 a) | / a

:: Matrix3 a = {a0 :: a, a1 :: a, a2 :: a, b0 :: a, b1 :: a, b2 :: a, c0 :: a, c1 :: a, c2 :: a}

Mat3determinant :: (Matrix3 a) -> a | *,-,+ a
Mat3Vec3Product :: (Matrix3 a) (Vector3 a) -> (Vector3 a) | *,+a
\end{CleanCode}

\subsection{ONB}
\begin{CleanCode}
:: ONB = {u_ :: (Vector3 Real), v_ :: (Vector3 Real), w_ :: (Vector3 Real), m_ :: (Matrix3 Real)}
setFromUW:: (Vector3 Real) (Vector3 Real) -> ONB
setFromV::(Vector3 Real) -> ONB
\end{CleanCode}

\subsection{Ray}
\begin{CleanCode}
:: Ray = {origin_ :: (Vector3 Real), direction_ :: (Vector3 Real)}
\end{CleanCode}

\subsection{IntersectionRecord}
\begin{CleanCode}
:: IntersectionRecord = {t_ :: Real, position_ :: (Vector3 Real),
normal_ :: (Vector3 Real), color_ :: (Vector3 Real)}
\end{CleanCode}

\newpage

\subsection{Triangle}
\begin{CleanCode}
:: Triangle = {colorT_ :: (Vector3 Real), a_ :: (Vector3 Real),
b_ :: (Vector3 Real), c_ :: (Vector3 Real)}
(intersect) :: Ray Triangle -> (Bool, IntersectionRecord)
\end{CleanCode}

\subsection{RayTriangleIntersector}
\begin{CleanCode}
IntersectRayTriangles :: Ray [Triangle] IntersectionRecord -> (Bool, IntersectionRecord)
\end{CleanCode}

\subsection{Buffer}
\begin{CleanCode}
InitBuff :: Int Int -> [[Vector3 Real]]
GetBuffElem :: [[Vector3 Real]] Int Int -> Vector3 Real
SetBuffElem :: [[Vector3 Real]] Int Int (Vector3 Real) -> [[Vector3 Real]]
SaveToPPM :: [[Vector3 Real]] String *World -> *World
LoadObj :: String *World -> ([Triangle], *World)
\end{CleanCode}

\subsection{PinHoleCamera}
\begin{CleanCode}
:: PinHoleCamera = { minX_ :: Real,
    maxX_ :: Real,
    minY_ :: Real,
    maxY_ :: Real,
    distance_ :: Real,
    resolution_ :: Vector2 Int,
    position_ :: Vector3 Real,
    up_ :: Vector3 Real,
    lookAt_ :: Vector3 Real,
    direction_ :: Vector3 Real,
    onb_ :: ONB }

initCamera :: Real Real Real Real Real (Vector2 Int) (Vector3 Real) (Vector3 Real) (Vector3 Real) 
-> PinHoleCamera
getWorldSpaceRay :: PinHoleCamera (Vector2 Int) -> Ray
\end{CleanCode}

\subsection{RayTracer}
\begin{CleanCode}
MAX_REAL =: 999999999999.9
resolution :: Vector2 Int
camera_position :: Vector3 Real
camera_up :: Vector3 Real
camera_lookAt :: Vector3 Real
camera :: PinHoleCamera
integrate :: (Vector2 Int) [Triangle] -> [[(Vector3 Real)]]
integrate2 :: Int Int [Triangle] -> Vector3 Real
\end{CleanCode}
\end{appendices}
\end{document}